\documentclass[conference]{IEEEtran}

\usepackage{amsmath}
\usepackage{amssymb}
\usepackage[dvips]{graphicx}
\usepackage{setspace}
\usepackage{epsfig}

\newcommand{\K}{{\sf{K}}}

\newcommand{\n}{\mathbf{n}}

\newcommand{\I}{\mathbf{I}}
\newcommand{\s}{\mathbf{s}}
\newcommand{\E}{\mathcal{E}}

\newcommand{\C}{{\sf{C}}}

\newcommand{\e}{\epsilon}

\newcommand{\rr}{{\mathbf{r}}}

\newcommand{\figsize}{0.34}

\newcommand{\tsnr}{{\text{\footnotesize{SNR}}}}

\newtheorem{theo:awgnderivatives}{Theorem}
\newtheorem{theo:cfderivatives}[theo:awgnderivatives]{Theorem}
\newtheorem{theo:ncfderivatives}[theo:awgnderivatives]{Theorem}
\newtheorem{theo:awgnderivfsk}{Proposition}
\newtheorem{theo:fskbitenergyasymp}[theo:awgnderivatives]{Theorem}
\newtheorem{theo:ncfskbitenergyasymp}[theo:awgnderivatives]{Theorem}
\newtheorem{theo:oofskasympt}[theo:awgnderivatives]{Theorem}

\newtheorem{corr:awgnasympt}{Corollary}
\newtheorem{corr:awgnbitenergy}[corr:awgnasympt]{Corollary}
\newtheorem{corr:awgnM3bitenergy}[corr:awgnasympt]{Corollary}
\newtheorem{corr:ncfadingasympt}[corr:awgnasympt]{Corollary}
\newtheorem{corr:ncfadingbitenergy}[corr:awgnasympt]{Corollary}
\newtheorem{corr:fadingfsk}[corr:awgnasympt]{Corollary}

\begin{document}

\title{On the Energy Efficiency of Orthogonal Signaling}

%\author{\authorblockN{Michael Shell} \and
%\and \authorblockN{James Kirk\\ and Montgomery Scott}
%\authorblockA{Starfleet Academy\\
%San Francisco, California 96678-2391\\ Telephone: (800)
%555--1212\\ Fax: (888) 555--1212}}

% avoiding spaces at the end of the author lines is not a problem with
% conference papers because we don't use \thanks or \IEEEmembership

\author{\authorblockN{Mustafa Cenk Gursoy}
\authorblockA{Department of Electrical Engineering\\
University of Nebraska-Lincoln, Lincoln, NE 68588\\ Email:
gursoy@engr.unl.edu}}

%\specialpapernotice{(Invited Paper)}

\maketitle

\begin{abstract}\footnote{This work was supported in part by the NSF
CAREER Grant CCF-0546384.} In this paper, transmission over the
additive white Gaussian noise (AWGN) channel, and coherent and
noncoherent fading channels using $M$-ary orthogonal frequency-shift
keying (FSK) or on-off frequency-shift keying (OOFSK) is considered.
The receiver is assumed to perform hard-decision detection. In this
setting, energy required to reliably send one bit of information is
investigated. It is shown that for fixed $M$ and duty cycle, bit
energy requirements grow without bound as the signal-to-noise ratio
($\tsnr$) vanishes. The minimum bit energy values are numerically
obtained for different values of $M$ and the duty cycle. The impact
of fading on the energy efficiency is identified. Requirements to
approach the minimum bit energy of $-1.59$ dB are determined.

%\emph{Index Terms:} Bit energy, spectral efficiency, AWGN channel,
%fading channels, phase-shift keying, frequency-shift keying, on-off
%keying, quantization, hard-decision detection.

\end{abstract}

\section{Introduction} \label{sec:intro}

Energy efficient transmission is of paramount importance in many
communication systems and particularly in mobile wireless systems
due to the scarcity of energy resources. Energy efficiency can be
measured by the energy required to send one information bit
reliably. It is well-known that for Gaussian channels subject to
average input power constraints, the minimum received bit energy
normalized by the noise spectral level is $\frac{E_b}{N_0}_{{\min}}
= -1.59$ dB regardless of the availability of channel side
information (CSI) at the
receiver (see e.g., \cite{Golay} -- %,\cite{Turin}, \cite{Jacobs},
%\cite{Pierce}, \cite{Lapidoth Shamai},
%\cite{Verdu_cost},
\cite{Verdu}). Golay \cite{Golay} showed that this
minimum bit energy can be achieved in the additive white Gaussian
noise (AWGN) channel by pulse position modulation (PPM) with
vanishing duty cycle when the receiver employs threshold detection.
Indeed, Turin \cite{Turin} proved that any orthogonal $M$-ary
modulation scheme with envelope detection at the receiver achieves
the normalized bit energy of $-1.59$ dB in the AWGN channel as $M
\to \infty$. It is further shown in \cite{Jacobs} and \cite{Pierce}
that $M$-ary orthogonal frequency-shift keying (FSK) achieves this
minimum bit energy asymptotically as $M \to \infty$ also in
noncoherent fading channels where neither the receiver nor the
transmitter knows the fading coefficients. As also well-known by now
in the digital communications literature \cite{Proakis}, these
results are shown by proving that the error probabilities of
orthogonal signaling can be made arbitrarily small as $M \to \infty$
as long as the normalized bit energy (or equivalently $\tsnr$ per
bit) is greater that $-1.59$ dB. These studies demonstrate that
asymptotically orthogonal signaling is optimally energy efficient
and highly resilient to fading even when the receiver performs
hard-decision detection. However, these asymptotical performances
require operation in the infinite bandwidth regime (when $M \to
\infty$) or signaling with unbounded peak-to-average power ratio,
i.e., vanishing duty cycle.
%As
%indicated by the unbounded growth of $M$, the minimum bit energy is
%in general achieved at infinite bandwidth or equivalently as the
%spectral efficiency (rate in bits per second divided by bandwidth in
%Hertz) goes to zero.

%As discussed in Section \ref{sec:intro}, orthogonal signaling is
%optimally energy efficient in the infinite bandwidth regime even if
%the receiver quantizes the received signals by performing
%hard-decision detection. For instance, PPM with vanishing duty cycle
%or $M$-ary FSK as $M \to \infty$ achieves the minimum bit energy of
%$-1.59$ dB.
In this paper, motivated by practical considerations and
limitations, we analyze the non-asymptotic energy efficiency of
orthogonal signaling. We consider $M$-ary FSK with finite $M$.
Additionally, we investigate the energy efficiency when FSK is
combined with on-off keying (or equivalently PPM) whose duty cycle
is small but nonzero. On-off FSK (OOFSK) modulation introduces
peakedness in both time and frequency, and provides improvements in
energy efficiency over on-off keying only or FSK only.

\vspace{-.27cm}
\section{Channel Model} \label{sec:channelmodel}

We consider the following channel model
\begin{gather}
\rr_k = h_k \s_{x_k} + \n_k \quad k = 1,2,3 \ldots
\end{gather}
where $x_k$ is the discrete input, $\s_{x_k}$ is the transmitted
signal when the input is $x_k$, and $\rr_k$ is the received signal
during the $k^{\text{th}}$ symbol duration. $h_k$ is the channel
gain. $h_k$ is a fixed constant in unfaded AWGN channels, while in
flat fading channels, $h_k$ denotes the fading coefficient.
$\{\n_k\}$ is a sequence of independent and identically distributed
(i.i.d.) zero-mean circularly symmetric Gaussian random vectors with
covariance matrix $E\{\n \n^\dagger\} = N_0 \I$ where $\I$ denotes
the identity matrix.
%The variance of $n_k$ is $E\{|n_k|^2\} = N_0$.
We assume that the system has an average energy constraint of
$E\{\|\s_{x_k}\|^2\} \ \le \E \quad \forall k$.

If $M$-ary orthogonal FSK modulation is used for transmission, then
$x_k \in \{1,2,\ldots, M\}$ and the transmitted signal has an $M$
complex-dimensional vector representation. If $x_k = m$,
$
\s_{x_k} = \s_m = (s_{m,1}, s_{m,2}, \ldots, s_{m,M})
$
where $s_{m,m} = \sqrt{\E}e^{j\theta_m}$ and $s_{m,i} = 0$ for all
$i \neq m$. The phases $\theta_m$ can be arbitrary. The received
signal $\rr_k$ and noise $\n_k$ are also $M$ dimensional. We assume
that the receiver quantizes the received vector $\rr_k$ by
performing energy detection.

\vspace{-.08cm}
\section{Energy Efficiency of FSK Modulation}
\label{sec:fsk}

In this section, we assume that the transmitter employs FSK
modulation and the receiver performs energy detection. In the
well-known noncoherent detection of FSK signals, $\s_i$ is declared
as the detected signal if the $i^{\text{th}}$ component of the
received vector $\rr$ has the largest energy, i.e.,
$%\begin{gather} \label{eq:fskdecisionrule}
|r_i|^2 > |r_j|^2 \quad \forall j \neq i. $ Note that this decision
rule is the maximum likelihood decision rule in AWGN, coherent
fading, and noncoherent Rician fading channels \cite{Lindsey},
\cite{gursoypeaky}, \cite{Proakis}. The output of the detector is
denoted by $y \in \{1,2,\ldots, M\}$. Note that with energy
detection, the channel can be now regarded as a symmetric discrete
channel with $M$ inputs and $M$ outputs.

Initially, we consider the AWGN channel and assume $h_k = 1 \,
\forall k$. The capacity which is achieved by equiprobable FSK
signals is given by\footnote{Throughout the paper, $\log$ is used to
denote the logarithm to the base $e$ i.e., the natural logarithm.}
\begin{gather}
C_M(\tsnr) = \log M + \sum_{l = 1}^{M} P_{l,1} \log P_{l,1}
\label{eq:mcapfsk}
\end{gather}
where $\tsnr = \frac{\E}{N_0}$ and $P_{l,1} = P(y = l | x = 1)$ is
the probability that $y = l$ given that $x = 1$. Using the results
on the error probabilities of noncoherent detection of FSK signals
(see e.g., \cite{Proakis}), we have
\begin{align}
P_{1,1} &= \sum_{n=0}^{M-1} \frac{(-1)^n}{n+1} \left(\!\!\!
\begin{array}{c}
M-1
\\
n
\end{array}\!\!\!\right)e^{-\frac{n}{n+1}\,\tsnr} \text{ and }
P_{l,1} = \frac{1-P_{1,1}}{M-1} \nonumber
%\label{eq:transitionprobfsk1}
\end{align}
for $l \neq 1$. Hence, the capacity can also be expressed as
%\cite{Stark}
\begin{gather}\label{eq:mcapfsk2}
\hspace{-.27cm}C_M(\tsnr) = \log M + P_{1,1} \log P_{1,1} +
(1-P_{11}) \log \frac{1-P_{1,1}}{M-1}.
\end{gather}
Next, we provide the behavior of the capacity in the low-$\tsnr$
regime.
\begin{theo:awgnderivfsk} \label{theo:awgnderivfsk}
The first derivative at zero $\tsnr$ of the capacity $C_M(\tsnr)$ in
(\ref{eq:mcapfsk}) is $ \dot{C}_M(0) = 0. $ Therefore, the bit
energy required at zero spectral efficiency is
\begin{gather}
\left.\frac{E_{b}}{N_0}\right|_{\C = 0} = \lim_{\tsnr \to 0}
\frac{\tsnr \log 2}{C_M(\tsnr)} =  \frac{\log 2}{\dot{C}_M(0)} =
\infty \quad \forall M \ge 2.
\end{gather}
\end{theo:awgnderivfsk}

\emph{Proof}: From (\ref{eq:mcapfsk2}), we can write
\begin{align}
\!\!\!\!\dot{C}_M(0) &= \dot{P}_{1,1}(\tsnr=0) (1+\log P_{1,1}(\tsnr
= 0)) - \dot{P}_{1,1}(\tsnr=0) \nonumber
\\
&- \dot{P}_{1,1}(\tsnr=0) \log
\frac{1-P_{1,1}(\tsnr=0)}{M-1}=0 \nonumber %\label{eq:cdotfsk1}
%\\
%&=(1+\log M)\left( \sum_{n=0}^{M-1} \frac{(-1)^n}{n+1} \left(\!\!\!
%\begin{array}{c}
%M-1
%\\
%n
%\end{array}\!\!\!\right)\left(\frac{-n}{n+1}\right) + \sum_{l = 2}^{M} \dot{P}_{l,1}|_{\tsnr=0}
%\right) \label{eq:cdotfsk3}
%\\
%&=(1+\log M)\left( \sum_{n=0}^{M-1} \frac{(-1)^n}{n+1} \left(\!\!\!
%\begin{array}{c}
%M-1
%\\
%n
%\end{array}\!\!\!\right)\left(\frac{-n}{n+1}\right) + \sum_{n=1}^{M-1} \frac{(-1)^{n+1}}{n+1} \left(\!\!\!
%\begin{array}{c}
%M-1
%\\
%n
%\end{array}\!\!\!\right)\left(\frac{-n}{n+1}\right)
%\right) \label{eq:cdotfsk4}
%\\
%&=0
\end{align}
Above, $\dot{P}_{1,1}(\tsnr=0)$ denotes the derivative of the
transition probability $P_{1,1}$ with respect to $\tsnr$ at $\tsnr =
0$ while $P_{1,1}(\tsnr = 0)$ is the value of the transition
probability at $\tsnr = 0$. The result is obtained by noting that
$P_{1,1}(\tsnr = 0) = 1/M$. \hfill $\square$

Although FSK is energy efficient asymptotically as $M \to \infty$,
Proposition \ref{theo:awgnderivfsk} shows that for finite $M$,
operating at very low $\tsnr$ levels is extremely energy inefficient
as the bit energy requirement increases without bound with
decreasing $\tsnr$ regardless of how large $M$ is. As a result, the
minimum bit energy is achieved at a nonzero spectral
efficiency\footnote{If FSK signals have a symbol duration of $T$,
the bandwidth requirement is $\frac{M}{T}$ and the spectral
efficiency is defined as $\C\left(\frac{E_b}{N_0} \right) =
\frac{\frac{C_M(\text{SNR})\log_2e}{T}}{\frac{M}{T}} =
\frac{C_M(\text{SNR})\log_2e}{M}$ bits/s/Hz.}, the value of which
can be found through numerical analysis. In \cite{Stark}, the
capacity and cutoff rate of $M$-ary FSK is studied in the AWGN,
coherent Rayleigh, and noncoherent Rician fading channels and it is
noted through numerical results that there exists an optimal code
rate for which the required bit energy is minimized. Here, we prove
that observation for any finite value of $M$. We note that energy
inefficiency at low $\tsnr$s is not a consequence of hard decision
detection. It is shown in \cite{gursoyoofsk} that the first
derivative of the FSK capacity is zero even if soft detection is
employed.

\begin{figure}
\begin{center}
\includegraphics[width = \figsize\textwidth]{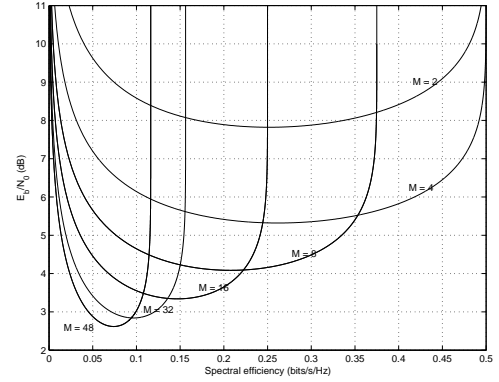}
\caption{Bit energy $E_b/N_0$ vs. Spectral efficiency $\C(E_b/N_0)$
for energy-detected $M$-ary FSK in the AWGN channel. $M =
2,4,8,16,32,48$.} \label{fig:fskawgn}
\end{center}
\end{figure}

Fig. \ref{fig:fskawgn} plots the bit energy $E_b/N_0$ curves as a
function of spectral efficiency for $M$-ary FSK in the AWGN channel
for different values of $M$.
In all cases, we observe that the minimum bit energy is achieved at
a nonzero spectral efficiency $\C^*$, and the bit energy
requirements increase to infinity as spectral efficiency decreases
to zero. Hence, operation below $\C^*$ should be avoided. Another
observation is that the minimum bit energy and the spectral
efficiency value at which the minimum is achieved decrease with
increasing $M$. For instance, when $M = 2$, the minimum bit energy
is 7.821 dB and is achieved at $\C^* = 0.251$ bits/s/Hz. On the
other hand, when $M = 48$, the minimum bit energy has decreased to
2.617 dB and is now attained at $\C^* = 0.074$ bits/s/Hz. On the
other hand, as $M$ increases, the minimum bit energy is achieved at
a higher $\tsnr$ value. Indeed, we can show that
\begin{align}
\hspace{-.4cm}\lim_{\substack{\e \to 0 \\ M \to \infty}}
\hspace{-.35cm}\left.\frac{C_M(\tsnr)}{\tsnr}\right|_{\tsnr =
(1+\e)\log M} \hspace{-.20cm}&= \lim_{\substack{\e \to 0 \\ M \to
\infty}} \frac{C_M((1+\e)\log M)}{(1+\e)\log M} \nonumber
\\
&= \lim_{\e \to 0}\frac{1}{1+\e} \,\lim_{M \to \infty} P_{1,1} = 1.
\label{eq:c/snrlim}
\end{align}
Hence, if $\tsnr$ grows logarithmically with increasing $M$, the bit
energy $\frac{E_b}{N_0} = \frac{\tsnr \log 2}{C_M(\tsnr)}$
approaches $\log 2 = -1.59$ dB. The proof of (\ref{eq:c/snrlim}) is
omitted because Turin \cite{Turin} has already shown that $-1.59$ dB
is achieved if the signal duration increases as $\log M$, which in
turn increases the $\tsnr$ logarithmically in $M$.

%The following result shows that as $M \to \infty$, the minimum bit
%energy approaches the fundamental limit of $-1.59$ dB.
%
%\begin{theo:fskbitenergyasymp} \label{theo:fskbitenergyasymp}
%As $M$ increases, the minimum bit energy required in the AWGN
%channel approaches
%\begin{gather}
%\lim_{M \to \infty} \, \frac{E_{b}}{N_0}_{\min} = \log_e2 = -1.59 \,
%dB
%\end{gather}
%Moreover, this minimum bit energy is achieved if the $\tsnr$ grows
%as $\log M$.
%\end{theo:fskbitenergyasymp}
%\emph{Proof}: See Appendix \ref{app:proofoftheo}.

Next, we consider fading channels. In coherent fading channels where
only the receiver has perfect knowledge of the fading coefficients,
the average capacity is
\begin{gather}
C_{M,c}(\tsnr) = \log M + \sum_{l = 1}^{M} E_h \{P_{l,1,h} \log
P_{l,1,h}\}
\end{gather}
where $ P_{1,1} = \sum_{n=0}^{M-1} \frac{(-1)^n}{n+1} \left(\!\!\!
\begin{array}{c}
M-1
\\
n
\end{array}\!\!\!\right)e^{-\frac{n}{n+1}|h|^2\tsnr}$ and $P_{l,1} = \frac{1-P_{1,1}}{M-1} \quad \text{for } l
\neq 1. $ We also consider noncoherent Rician fading channels where
neither the transmitter nor the receiver knows the fading
coefficients. In this case, we assume that $\{h_k\}$ are i.i.d.
complex Gaussian random variables with mean $E\{h_k\} = d$ and
variance $E\{|h_k - d|^2\} = \gamma^2$. In the noncoherent Rician
fading channel, the capacity of $M$-ary FSK modulation is
\begin{gather}
C_{M,nc}(\tsnr) = \log M + \sum_{l = 1}^{M} P_{l,1} \log P_{l,1}
\end{gather}
where $ P_{1,1} = \sum_{n=0}^{M-1} \frac{(-1)^n}{n(1+ \gamma^2
\tsnr)+1} \left(\!\!\!
\begin{array}{c}
M-1
\\
n
\end{array}\!\!\!\right)e^{-\frac{n}{n(1+\gamma^2
\tsnr)+1}|d|^2\tsnr}$ and $P_{l,1} = \frac{1-P_{1,1}}{M-1} \quad
\text{for } l \neq 1. \label{eq:transitionprobfsknoncoh1} $ We note
that these transition probabilities are obtained from the error
probability expressions of FSK modulation in noncoherent Rician
fading channels (see e.g., \cite{Lindsey}, \cite{gursoypeaky}).
Since the presence of fading unknown at the transmitter only
decreases the performance, we immediately have the following
corollary to Proposition \ref{theo:awgnderivfsk}.
\begin{corr:fadingfsk}
The first derivatives at zero $\tsnr$ of the capacities
$C_{M,c}(\tsnr)$ and $C_{M,nc}(\tsnr)$ are equal to zero, i.e.,
$\dot{C}_{M,c}(0) = \dot{C}_{M,nc}(0) = 0$. Therefore, the bit
energy required at zero spectral efficiency is infinite in both
coherent and noncoherent fading channels, i.e., $
\left.\frac{E_{b,c}}{N_0}\right|_{\C = 0} =
\left.\frac{E_{b,nc}}{N_0}\right|_{\C = 0} = \infty. $
\end{corr:fadingfsk}

\begin{figure}
\begin{center}
\includegraphics[width = \figsize\textwidth]{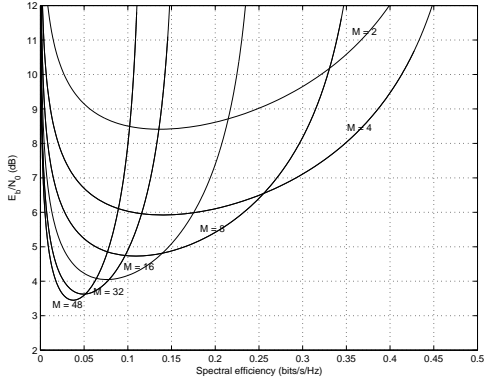}
\caption{Bit energy $E_b/N_0$ vs. Spectral efficiency $\C(E_b/N_0)$
for energy-detected $M$-ary FSK in the coherent Rician fading
channel with Rician factor $\K =1$.} \label{fig:fskcRicianK1}
\end{center}
\end{figure}

\begin{figure}
\begin{center}
\includegraphics[width = \figsize\textwidth]{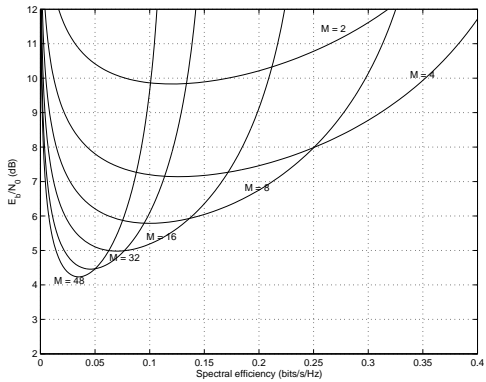}
\caption{Bit energy $E_b/N_0$ vs. Spectral efficiency $\C(E_b/N_0)$
for energy-detected $M$-ary FSK in the noncoherent Rician fading
channel with Rician factor $\K =1$.} \label{fig:fskncRicianK1}
\end{center}
\end{figure}

\begin{figure}
\begin{center}
\includegraphics[width = \figsize\textwidth]{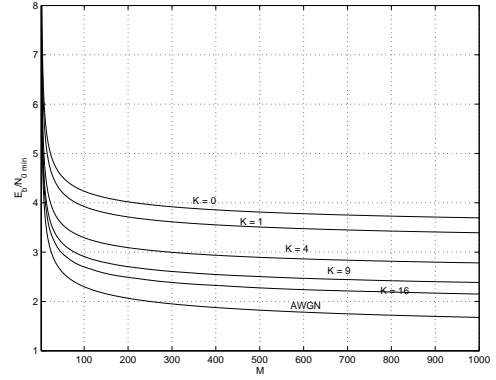}
\caption{Minimum bit energy ${E_b/N_0}_{\min}$ vs. $M$ for $M$-ary
FSK in the AWGN channel and noncoherent Rician fading channels with
Rician factors $\K = 0,1,4,9,16$. Note that $K=0$ corresponds to the
noncoherent Rayleigh channel.} \label{fig:minbitenergy_vs_M2}
\end{center}
\end{figure}

Figures \ref{fig:fskcRicianK1} and \ref{fig:fskncRicianK1} plot the
bit energy curves for $M$-ary FSK transmission over coherent and
noncoherent Rician fading channels.
As predicted, the bit energy levels for all values of $M$ increase
without bound as the spectral efficiency decreases to zero. Due to
the presence of fading, the minimum bit energies have increased with
respect to those achieved in the AWGN channel. For instance, when $M
= 48$, the minimum bit energies are now $E_b/N{_0}_{\min} = 3.45$ dB
in the coherent Rician fading channel and $E_b/N{_0}_{\min} = 4.23$
dB in the noncoherent Rician fading channel. We again observe that
the minimum bit energy decreases with increasing $M$. Fig.
\ref{fig:minbitenergy_vs_M2} provides the minimum bit energy values
as a function of $M$ in the AWGN and noncoherent Rician fading
channels with different Rician factors. In all cases, the minimum
bit energy decreases with increasing $M$. However, Fig.
\ref{fig:minbitenergy_vs_M2} indicates that approaching $-1.59$ dB
is very slow and demanding in $M$.
\begin{figure}
\begin{center}
\includegraphics[width = \figsize\textwidth]{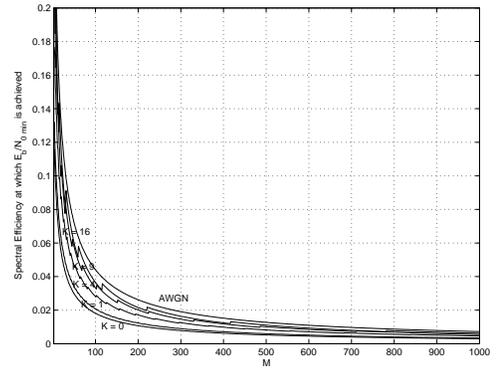}
\caption{Spectral efficiency at which ${E_b/N_0}_{\min}$ is achieved
vs. $M$ for $M$-ary FSK in the AWGN channel and noncoherent Rician
fading channels with Rician factors $\K = 0,1,4,9,16$.}
\label{fig:spectraleff_vs_M}
\end{center}
\end{figure}
\begin{figure}
\begin{center}
\includegraphics[width = \figsize\textwidth]{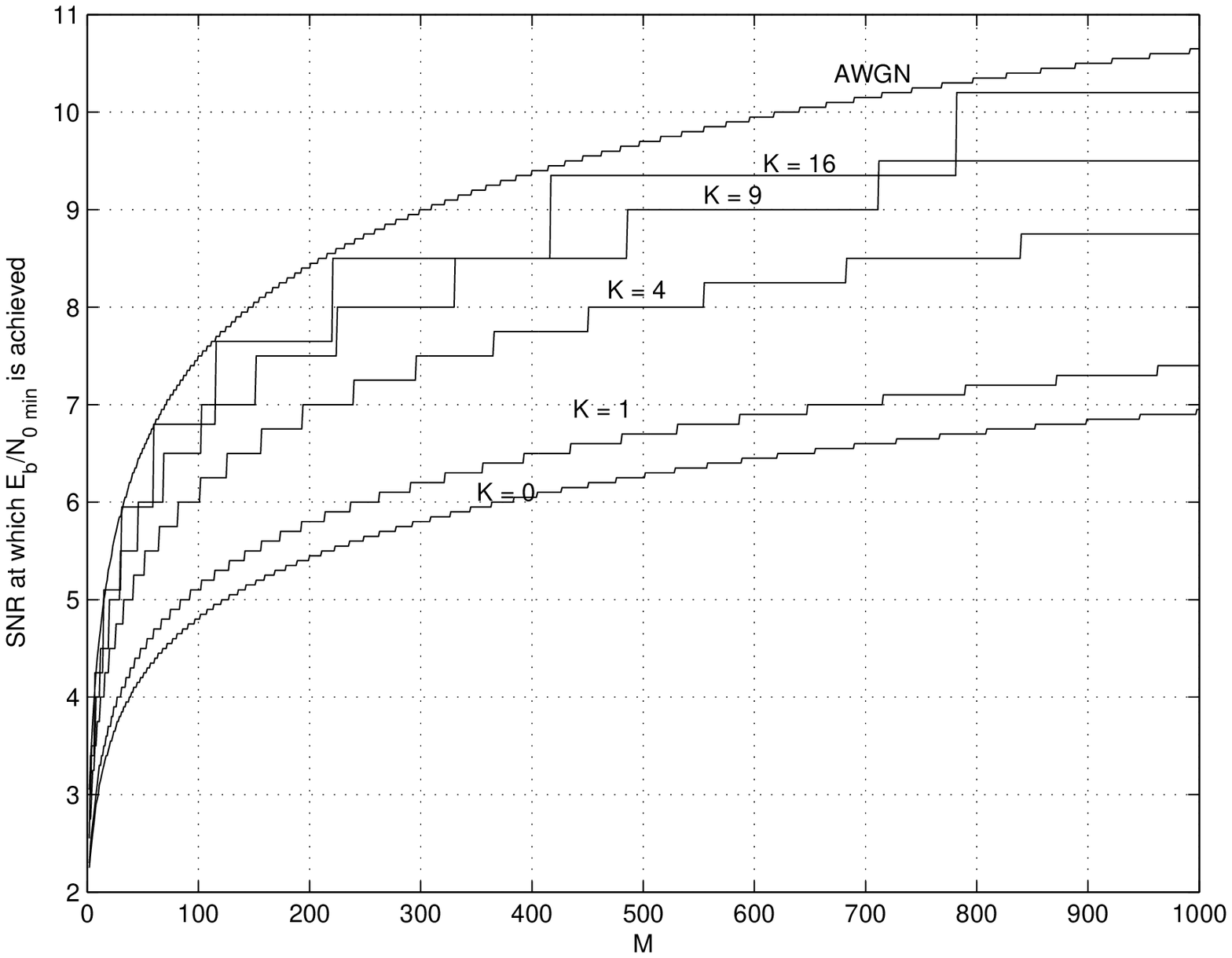}
\caption{$\tsnr$ at which ${E_b/N_0}_{\min}$ is achieved vs. $M$ for
$M$-ary FSK in the AWGN channel and noncoherent Rician fading
channels with Rician factors $\K = 0,1,4,9,16$.}
\label{fig:snr_vs_M}
\end{center}
\end{figure}
 In this figure, we also note the
energy penalty due to the presence of unknown fading. But, as the
Rician factor $\K$ increases, the noncoherent Rician channel
approaches to the AWGN channel and so do the minimum bit energy
requirements. Figures \ref{fig:spectraleff_vs_M} and
\ref{fig:snr_vs_M} plot the spectral efficiencies and average
received $\tsnr$ values at which $E_b/N{_0}_{\min}$ is achieved as a
function of $M$. As also observed in Figs. \ref{fig:fskawgn} and
\ref{fig:fskncRicianK1}, $E_b/N{_0}_{\min}$ is achieved at lower
spectral efficiencies as $M$ increases. From Fig.
\ref{fig:spectraleff_vs_M}, we note that the required spectral
efficiencies are lower and hence the bandwidths requirements are
higher in noncoherent fading channels. In Fig. \ref{fig:snr_vs_M},
we see that the $\tsnr$ levels at which $E_b/N{_0}_{\min}$ is
achieved increases with increasing $M$. As predicted in
(\ref{eq:c/snrlim}), the increase is logarithmic in $M$ in the AWGN
channel. Similar rates of increase are also noted for noncoherent
fading channels.

%Indeed, using similar approaches as in the proof of Theorem
%\ref{theo:fskbitenergyasymp}, we can show the following result.
%
%\begin{theo:ncfskbitenergyasymp} \label{theo:ncfskbitenergyasymp}
%As $M$ increases, the minimum bit energy required in the noncoherent
%Rician fading channels channel is asymptotically upper bounded by
%\begin{gather}
%\lim_{M \to \infty} \, \frac{E_{b}}{N_0}_{\min} \le
%\frac{\log_e2}{Q_1\left( \sqrt{\frac{2|d|^2}{\gamma^2}},
%\sqrt{\frac{2}{\gamma^2}}\right)} \label{eq:asymbiterrorupper}
%\end{gather}
%where $Q_1(\cdot,\cdot)$ is the Marcum $Q$-function \cite{Simon}.
%\end{theo:ncfskbitenergyasymp}
%\emph{Proof}: See Appendix \ref{app:ncproofoftheo}
%
%If $|d|^2 = 5$ and $\gamma^2 = 1$, hence the Rician factor is $\K =
%5$, the upper bound in (\ref{eq:asymbiterrorupper}) is $-1.49$ dB
%while if $|d|^2 = 10$, the upper bound $-1.5892$ dB. Hence, we are
%very close to $-1.59$ dB even for these relatively small Rician
%factor values. On the other hand, if $|d|^2 = 0$ and we have the
%noncoherent Rayleigh fading channel, the left hand side of
%(\ref{eq:asymbiterrorupper}) is $2.75$ dB.

\section{Energy Efficiency of OOFSK Modulation}

In the previous section, we have seen that the minimum bit energy
decreases as the number of orthogonal frequencies, $M$, increases.
Another technique to improve the energy efficiency is to increase
the peakedness of the signal in the time domain. In
\cite{gursoyoofsk}, we have introduced the on-off frequency-shift
keying (OOFSK) modulation by overlaying frequency-shift keying on
on-off keying. In $M$-ary OOFSK modulation, the transmitter either
sends no signal with probability $1-\nu$ or sends one of $M$
orthogonal FSK signals each with probability $\nu/M$. In this
setting,  $\nu \in (0,1]$ can be seen as the duty cycle of the
transmission. While the FSK signals have energy $\E/\nu$, the
average energy of OOFSK modulation is $\E$. Hence, the
peak-to-average power ratio of OOFSK is $1/\nu$. No transmission is
denoted by $\s_0 = (0,0,\ldots,0)$. The FSK signals have the
following geometric representations:
\begin{gather}
\s_m = (s_{m,1}, s_{m,2}, \ldots, s_{m,M}) \quad m = 1,2,\ldots, M,
\end{gather}
where $s_{m,m} = \sqrt{\E/\nu}\,e^{j\theta_m}$ and $s_{m,i} = 0$ for
$i \neq m$. Note that in $M$-ary OOFSK modulation, we have $M+1$
possible input signals including the no signal transmission.
%Choosing $M=1$ reduces this signaling to pure on-off keying.
We again assume that the received signal is  hard-decision detected
at the receiver. In \cite{gursoypeaky}, maximum a posteriori
probability (MAP) detection rule for OOFSK modulation  is identified
and the error probability expressions are obtained. The optimal
detection rule is given by the following: $\s_i$ for $i \neq 0$ is
detected if $|r_i|^2 > |r_j|^2$ $\forall j \neq i$ and $|r_i|^2 >
\tau$ where $\tau = \left\{
\begin{array}{ll}
\frac{\left[I_0^{-1}\left(\xi\right)\right]^2}{4\alpha^2} & \xi \ge
1
\\
0 & \xi < 1
\end{array}\right.$, $ \xi=\frac{M(1-\nu)\,e^{\alpha^2}}{\nu}$, and $\alpha^2 = \frac{\tsnr}{\nu}$. No transmission and hence $\s_0$ is detected if
$|r_i|^2 < \tau$ $\forall i$. After hard-decision detection, the
channel can be regarded as a discrete channel with $M+1$ inputs and
$M+1$ outputs. From the error probability analysis in
\cite{gursoypeaky}, we have the following expressions for the
transition probabilities in the AWGN channel:
\begin{gather}
P_{0,0} = (1-e^{-\tau})^M, \quad  P_{l,0} =
\frac{1}{M}(1-(1-e^{-\tau})^M) \label{eq:transitionproboofsk2}
\\
\hspace{-1cm}P_{l,l} =
\sum_{n=0}^{M-1}\frac{(-1)^n}{n+1}\left(\begin{array}{cc}M-1\\n\end{array}\right)
e^{-\frac{n}{n+1}\alpha^2}
Q_1\left(\sqrt{\frac{2}{n+1}}\,\alpha,\sqrt{2(n+1)\tau}\right)
\label{eq:transitionproboofsk3}
\\
P_{0,l} = (1-e^{-\tau})^{M-1}\left(1-Q_1\left( \sqrt{2}\,\alpha,
\sqrt{2\tau} \right)\right) \label{eq:transitionproboofsk4}
\end{gather}
for $l= 1,2,\ldots,M$, and
\begin{gather}
\!\!\!\!\!\!P_{l,m} = \frac{1}{M-1}\left(1-P_{m,m}-P_{0,m}\right)
\forall l,m\neq 0, \text{and } l \neq m.
\label{eq:transitionproboofsk5}
\end{gather}
In the above expressions, $Q_1(\cdot,\cdot)$ is the Marcum
$Q$-function \cite{Simon}, and $I_0^{-1}$ is the functional inverse
of the zeroth order modified Bessel function of the first kind. The
rate achieved by the $M$-ary OOFSK modulation with duty cycle $\nu$
and equiprobable FSK signals is
\begin{align}
I_{M}(\tsnr,\nu) &= H(y) - H(y|x)
\\
&=-\left((1-\nu)P_{0,0} + \nu P_{0,1} \right) \log
\left((1-\nu)P_{0,0} + \nu P_{0,1} \right) \nonumber
\\
&\,\,\,\,\,\,-M\left( (1-\nu)P_{1,0} + \frac{\nu}{M} P_{1,1} +
\frac{(M-1)\nu}{M} P_{1,2}\right) \nonumber
\\
&\,\,\,\,\,\,\,\,\,\,\,\,\,\times\log \left( (1-\nu)P_{1,0} +
\frac{\nu}{M} P_{1,1} + \frac{(M-1)\nu}{M} P_{1,2}\right) \nonumber
\\
&\,\,\,\,\,\,+ (1-\nu) \left( P_{0,0} \log P_{0,0} + M P_{1,0} \log
P_{1,0} \right) \nonumber
\\
&\,\,\,\,\,\,+\nu \left( P_{0,1} \log P_{0,1} + P_{1,1} \log P_{1,1}
+ (M-1) P_{2,1} \log P_{2,1}\right). \label{eq:achievableoofsk}
\end{align}
It is shown in \cite{gursoyoofsk} that in the AWGN channel, the
capacity with soft detection of OOFSK signaling has a zero slope at
$\tsnr = 0$. Since hard-decision detection decreases the capacity,
we can immediately conclude that $\dot{I}_{M}(0,\nu) = 0$ for any
fixed nonzero value of $\nu$, and hence the bit energy required at
zero spectral efficiency is still infinite. On the other hand, we
know from \cite{Golay} and \cite{Verdu} that if the duty cycle $\nu$
vanishes simultaneously with $\tsnr$, the minimum bit energy of
$-1.59$ dB can be achieved. The following result identifies the rate
at which $\nu$ should decrease as $\tsnr$ gets smaller.

\begin{theo:oofskasympt} \label{theo:oofskasympt}
Assume that $\nu = \frac{\tsnr}{(1+\e) \log \frac{1}{\tsnr}}$ for
$\tsnr < 1$ and for some $\e > 0$. Then, we have
\begin{gather}
\lim_{\e \to 0} \lim_{\tsnr \to 0} \frac{I_M(\tsnr,\nu)}{\tsnr} = 1
\intertext{ and hence } \lim_{\e \to 0} \lim_{\tsnr \to 0}
\frac{\tsnr \log 2}{I_M(\tsnr,\nu)} = \log 2 = -1.59 \text{ dB}.
\end{gather}
\end{theo:oofskasympt}

\emph{Proof:} Note that as $\tsnr \to 0$, $\nu \to 0$ and $\alpha^2
= \frac{\tsnr}{\nu} = (1+\e) \log \frac{1}{\tsnr} \to \infty$. It
can also be seen that $\xi \to \infty$ and $\tau \to \infty$ as
$\tsnr$ diminishes. From (\ref{eq:transitionproboofsk2}), we
immediately note that $P_{0.0} \to 1$ and $P_{l,0} \to 0$ for $l =
2,\ldots,M$. In (\ref{eq:transitionproboofsk3}), all the terms in
the summation other than for $n = 0$ vanishes because $\alpha^2 \to
\infty$. Therefore, in order for $P_{l,l}$ for $l=1,\ldots,M$ to
approach 1, we need $Q_1(\sqrt{2}\, \alpha, \sqrt{2\tau}) \to 1$.
Also note that if $Q_1(\sqrt{2}\, \alpha, \sqrt{2\tau}) \to 1$, then
we can observe from (\ref{eq:transitionproboofsk4}) and
(\ref{eq:transitionproboofsk5}) that $P_{0,l} \to 0$ and $P_{l,m}
\to 0$. Hence, eventually all crossover error probabilities will
vanish and correct detection probabilities will be 1.

In \cite{Simon}, it is shown that $Q_1(a, a \zeta) \ge 1 -
\frac{\zeta}{1-\zeta} e^{-\frac{a^2 (1-\zeta)^2}{2}} \quad 0 \le
\zeta < 1. $ From this lower bound we can immediately see that $
\lim_{\tsnr \to 0} Q_1(\sqrt{2}\, \alpha, \sqrt{2\tau}) = 1$ if
$\lim_{\tsnr \to 0} \frac{\tau}{\alpha^2} < 1.$ Note that both
$\alpha^2$ and $\tau$ grow without bound as $\tsnr \to 0$. Recall
that $\tau =
\frac{\left[I_0^{-1}\left(\xi\right)\right]^2}{4\alpha^2}$.
Equivalently, we have $I_0(\sqrt{4\alpha^2 \tau}) = \xi =
\frac{M(1-\nu)\,e^{\alpha^2}}{\nu}$. Using the asymptotic form
$I_0(x) = \frac{1}{\sqrt{2\pi x}}\, e^x +
\mathcal{O}\left(\frac{1}{x^{3/2}}\right)$ \cite{Butkov} for large
$x$, we can easily show that $\lim_{\tsnr \to 0}
\frac{\tau}{\alpha^2} = \left(\frac{1+\e/2}{1+\e}\right)^2 < 1$
$\forall \e > 0$ if $\nu = \frac{\tsnr}{(1+\e) \log
\frac{1}{\tsnr}}$. Therefore, if $\nu$ decays at this rate, the
error probabilities go to zero. It can then be shown that
$\lim_{\tsnr \to 0} \frac{I_M(\tsnr,\nu)}{\tsnr} = \frac{1}{1+\e}$.
Since results hold for any $\e > 0$, letting $\e \to 0$ gives the
desired result. \hfill $\square$

We note that Zheng \emph{et al.} have shown in \cite{Zheng2} that
the low $\tsnr$ capacity of unknown Rayleigh fading channel can be
approached by on-off keying if $\frac{\log \frac{1}{\tsnr}}{\log
\log \frac{1}{\tsnr}} \le \alpha^2 \le \log \frac{1}{\tsnr}$. We see
a similar behavior here when FSK signals are sent over the AWGN
channel and envelope detected.

In coherent fading channels where the receiver has perfect knowledge
of the fading coefficients, the transition probabilities are the
same as those in
(\ref{eq:transitionproboofsk2})-(\ref{eq:transitionproboofsk5}) with
the only difference that we now have $\alpha^2 =
\frac{\tsnr}{\nu}|h|^2$. As a result, the achievable rates
$I_{M}(\tsnr,\nu,|h|^2)$ are also dependent on the fading
coefficients and average achievable rates are obtained by finding
the expected value $E_{|h|^2} \{I_{M}(\tsnr,\nu,|h|^2)\}$.

In noncoherent Rician fading channels, the transition probabilities
\cite{gursoypeaky} are
\begin{gather}
P_{0,0} = (1-e^{-\tau})^M, \quad P_{l,0} =
\frac{1}{M}(1-(1-e^{-\tau})^M)\label{eq:transitionproboofsknc2}
\\
\!\!\!\!\!\!\!\!\!\!\!\!\!\!\!P_{l,l} =
\sum_{n=0}^{M-1}(-1)^n\left(\!\!\!\begin{array}{cc}M-1\\n\end{array}\!\!\!\right)
\frac{e^{-\frac{n \alpha^2
|d|^2}{n(1+\gamma^2\alpha^2)+1}}}{n(1+\gamma^2\alpha^2 )+1}
\nonumber
\\
\hspace{0.5cm}\times Q_1\left(\sqrt{\frac{2 \alpha^2
|d|^2}{(1+\gamma^2\alpha^2)(n(1+\gamma^2\alpha^2)+1)}},\sqrt{\frac{2(n(1+\gamma^2\alpha^2)+1)\tau}{(1+\gamma^2\alpha^2)}}\right)
\nonumber \label{eq:transitionproboofsknc3}
\\
P_{0,l} = (1-e^{-\tau})^{M-1}\left(1-Q_1\left( \sqrt{\frac{2\alpha^2
|d|^2}{1+\gamma^2 \alpha^2}}, \sqrt{\frac{2\tau}{1+\gamma^2
\alpha^2}} \right)\right)\label{eq:transitionproboofsknc4}
\end{gather}
for $l= 1,2,\ldots,M$, and
\begin{gather}
\!\!\!\!\!\!\!P_{l,m} = \frac{1}{M-1}\left(1-P_{m,m}-P_{0,m}\right)
\forall l,m\neq 0, \text{and } l \neq m.
\label{eq:transitionproboofsknc5}
\end{gather}
In the above expressions, $\tau =\left\{
\begin{array}{ll}
 \Phi^{-1}(\xi) & \xi \ge 1
 \\
 0 & \xi < 1
\end{array}\right.$, $\Phi(x) = e^{\frac{\alpha^2 \gamma^2 x}{1 + \alpha^2 \gamma^2}}
I_0\left( \frac{2\sqrt{x \, \alpha^2 |d|^2}}{1 + \alpha^2
\gamma^2}\right), \text{ and } \xi = \frac{M(1-\nu)}{\nu} (1+
\alpha^2 \gamma^2) \, e^{\frac{\alpha^2 |d|^2}{1 + \alpha^2
\gamma^2}}.$ The achievable rates can be obtained from
(\ref{eq:achievableoofsk}). If $|d| \ge 1$, then following the same
steps as in the proof of Theorem \ref{theo:oofskasympt}, we can show
that the minimum bit energy of $-1.59$ dB is achieved as $\tsnr \to
0$ if $\nu = \frac{\tsnr}{(1+\e) \log \frac{1}{\tsnr}}$.

\begin{figure}
\begin{center}
\includegraphics[width = \figsize\textwidth]{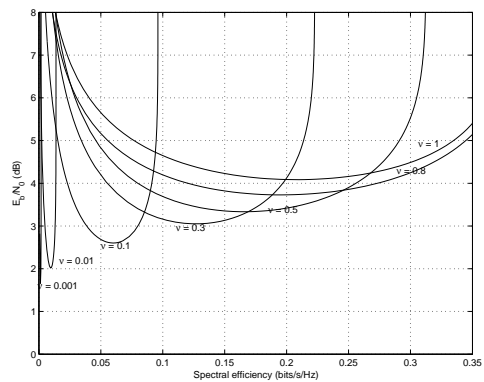}
\caption{Bit energy $E_b/N_0$ vs. Spectral efficiency $\C(E_b/N_0)$
for 8-OOFSK in the AWGN channel. The duty cycle values are $\nu =
1,0.8,0.5,0.3,0,1,0.01$ and 0.001.} \label{fig:oofskawgn}
\end{center}
\end{figure}

\begin{figure}
\begin{center}
\includegraphics[width = \figsize\textwidth]{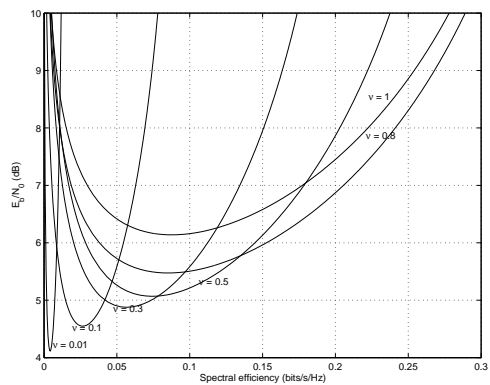}
\caption{Bit energy $E_b/N_0$ vs. Spectral efficiency $\C(E_b/N_0)$
for  8-OOFSK in the noncoherent Rayleigh fading channel. The duty
cycle values are $\nu = 1,0.8,0.5,0.3,0,1,0.01$ and 0.001.}
\label{fig:oofskncRayleigh}
\end{center}
\end{figure}

Figs. \ref{fig:oofskawgn} and \ref{fig:oofskncRayleigh} plot the bit
energies as a function of spectral efficiency of 8-OOFSK with
different duty cycle factors in the AWGN and noncoherent Rayleigh
fading channels.
We immediately observe that decreasing the duty cycle lowers the
minimum bit energy. Hence, increasing the signal peakedness in the
time domain improves the energy efficiency. In the AWGN channel,
while regular 8-FSK (8-OOFSK with $\nu =1$) has $E_b/N{_0}_{\min}=
4.08$ dB, 8-OOFSK with $\nu =0.01$ has $E_b/N{_0}_{\min} = 2.017$
dB. However, this energy gain is obtained at the cost of increased
peak-to-average ratio. We also note that unknown fading again
induces an energy penalty with respect to the AWGN channel as
observed by comparing Figs. \ref{fig:oofskawgn} and
\ref{fig:oofskncRayleigh}.

%\section{Conclusion} \label{sec:conclusion}
%
%In this paper, we have analyzed the impact of quantization on the
%energy efficiency of phase modulation and frequency modulation
%together with on-off keying. We have considered a system in which
%the receiver quantizes the received signals by perfoming
%hard-decision detection. By analyzing general $M$-ary modulation
%schemes, we have addressed the impact of having different
%quantization levels. We have also considered energy detected $M$-ary
%FSK transmission over the AWGN and fading channels. We have shown
%that bit energy requirements grow without bound as $\tsnr$ vanishes.
%Through numerical results, we have investigated the value of the
%minimum bit energy for different values of $M$ in different
%channels. We have finally considered OOFSK modulation. For a fixed
%value of duty cycle, we have shown that the bit energy requirements
%still grow without bound as $\tsnr$ decreases. On the other hand, we
%have shown through numerical results that the minimum bit energy
%decreases with decreasing duty cycle. We have proved that if the
%duty cycle decreases as $\frac{\tsnr}{\log \frac{1}{\tsnr}}$, the
%minimum bit energy of
%$-1.59$ dB can be approached. %in the AWGN channel.

%\vspace{-cm}

%\begin{figure}
%\begin{center}
%\includegraphics[width = 0.65\textwidth]{fskcRayleigh.eps}
%\caption{Bit energy $E_b/N_0$ vs. Spectral efficiency $\C(E_b/N_0)$
%for $M$-ary FSK with a hard-decision detection in the coherent
%Rayleigh fading channel.} \label{fig:fskcRayleigh}
%\end{center}
%\end{figure}


\begin{thebibliography}{}

%\bibitem{Shannon} C. E. Shannon, ``Communication in the presence of noise,
%" \emph{Proc. IRE}, vol.~37, pp. 10-21, Jan. 1949.

\bibitem{Golay} M. J. E. Golay, ``Note on the theoretical efficiency of information reception
with PPM," \emph{Proc. IRE}, vol.~37, pp.~1031, Sept. 1949.

\bibitem{Turin} G. L. Turin, ``The asymptotic behavior of ideal $M$-ary systems,"
\emph{Proc. IRE}, vol.~47, pp.~93-94, Jan. 1959.

\bibitem{Jacobs} I. Jacobs, ``The asymptotic behavior of incoherent $M$-ary communications systems,"
\emph{Proc. IEEE}, vol.~51, pp.~251-252, Jan. 1963.

\bibitem{Pierce} J. N. Pierce, ``Ultimate performance of $M$-ary transmissions on fading channels," \emph{IEEE Trans. Inform. Theory},
vol.~IT-12, pp.~2-5, Jan. 1966.

\bibitem{Lapidoth Shamai} A. Lapidoth and S. Shamai (Shitz), ``Fading channels: How perfect need `perfect side
information' be?," \emph{IEEE Trans. Inform. Theory}, vol.~48,
pp.~1118-1134, May 2002.

\bibitem{Verdu_cost} S. Verd\'u, ``On channel capacity per unit cost," \emph{IEEE Trans. Inform.
Theory}, vol.~36, pp.~1019-1030, Sep. 1990

%\bibitem{Pierce_phase} J. R. Pierce, ``Comparison of three-phase
%modulation with two-phase and four-phase modulation," \emph{IEEE
%Trans. Commun}, vol.~28, pp.~1098-1099, July 1980.

\bibitem{Verdu} S. Verd\'u, ``Spectral efficiency in the wideband regime," \emph{IEEE
Trans. Inform. Theory}, vol.~48, pp.~1319-1343, June 2002.


%\bibitem{Kramer} G. Kramer, A. Ashikhmin, A. J. van Wijngaarden, and X. Wei, ``Spectral efficiency of coded
%phase-shift keying for fiber-optic communication," \emph{IEEE/OSA J.
%Lightwave Technol.}, vol.~21, pp.~2438-2445, Oct. 2003.
%
%\bibitem{Wyner} A. D. Wyner, ``Bounds on communication with
%polyphase coding," \emph{Bell Syst. Tech. J.,} vol. XLV, pp.
%523-559, Apr. 1966.
%
%\bibitem{Kaplan} G. Kaplan and S. Shamai (Shitz), ``On the
%achievable information rates of DPSK," \emph{IEE Proceesings}, vol.
%139, pp. 311-318, June 1992.
%
%\bibitem{Peleg} M. Peleg and Shlomo Shamai (Shitz), ``On the
%capacity of the blockwise incoherent MPSK channel," \emph{IEEE
%Trans. Commun}, vol.~46, pp.~603-609, May 1998.
%
%\bibitem{Gursoy-part2} M. C. Gursoy, H. V. Poor, and S. Verd\'u, ``The noncoherent Rician fading
%channel -- Part II : Spectral efficiency in the low power regime,"
%\emph{IEEE Trans. Wireless Commun.}, vol. 4, no. 5, pp. 2207-2221,
%Sept. 2005.
%
%\bibitem{Zhang} W. Zhang and J. N. Laneman,``How good is phase-shift keying for
%peak-limited Rayleigh fading channels in the low-SNR regime?,"
%\emph{IEEE Trans. Inform. Theory}, vol.~53, pp.~236 - 251, Jan.
%2007.
%
%\bibitem{Luo} X. Luo and G. B. Giannakis, ``Energy-constrained optimal
%quantization for wireless sensor networks," IEEE SECON, pp. 272-278,
%4-7 Oct. 2004.

\bibitem{Simon} M. K. Simon and M.-S. Alouni,
 ''A unified approach to the performance analysis of digital
 communication over generalized fading channels", \emph{Proc. of the IEEE}, vol. 86, no. 9, pp. 1860-1877, Sept. 1998

\bibitem{Lindsey} W. C. Lindsey, ``Error probabilities for Rician fading multichannel reception of binary and N-ary Signals",
\emph{IEEE Trans. Inform. Theory}, vol.10, pp. 339-350, Oct. 1964.

\bibitem{Stark} W. E. Stark, ``Capacity and cutoff rate of
noncoherent FSK with nonselective Rician fading," \emph{IEEE Trans.
Commun}, vol.~33, pp.~1153-1159, Nov. 1985.

\bibitem{Zheng2} L. Zheng, D. N. C. Tse, and M. M\'edard ``Channel coherence in the low
SNR regime," \emph{IEEE Trans. Inform. Theory}, vol.~53,
pp.~976-997, March 2007.

%\bibitem{wang} Q. Wang and M.C. Gursoy, ``Error Performance of OOFSK Signaling over
%Fading Channels," Proceedings of the 40th Annual Conference on
%Information Sciences and Systems, Princeton University, Princeton,
%NJ, March, 22-24, 2006.

\bibitem{gursoyoofsk} M. C. Gursoy, H. V. Poor, S.
 Verd\'u, ``On-Off frequency-shift keying for wideband fading channels"
 \emph{EURASIP Journal on Wireless Communications and
 Networking}, 2006.

\bibitem{gursoypeaky} M. C. Gursoy, ``Error rate analysis for peaky
signaling over fading channels," submitted to the IEEE Transactions
on Communications, 2007 (available at
http://www.ee.unl.edu/faculty-staff/gursoy.shtml).

\bibitem{Proakis} J. G. Proakis, \emph{Digital Communications.}
New York: McGraw-Hill, 1995.

%\bibitem{ProakisSalehi} J. G. Proakis and M. Salehi, \emph{Fundamentals of Communication Systems.}
%New York: Pearson Prentice Hall, 2005.

\bibitem{Cover} T. M. Cover and J. A. Thomas, \emph{Elements of Information
Theory.} New York: Wiley, 1991.

\bibitem{Butkov} E. Butkov, \emph{Mathematical Physics.} Addison-Wesley, 1968.

\end{thebibliography}
\end{document}